\documentclass[%
twocolumn,showpacs,
amsmath,prd,amssymb,
showkeys,superscriptaddress,
longbibliography, nofootinbib
]{revtex4-2}

\usepackage{graphicx}% Include figure files
\usepackage{dcolumn}% Align table columns on decimal point
\usepackage{multirow}
\usepackage{mathtools}
\usepackage{amsmath}
\usepackage{scalerel}
\usepackage{bm}% bold math
\usepackage{mathrsfs}
 \usepackage{color}

\usepackage{natbib}
\usepackage{multirow}

  \usepackage{array}
  \usepackage{verbatim}
  \usepackage{amstext}
  \usepackage{amsmath}
  \usepackage{amssymb}
  \usepackage{slashed}
  \usepackage{indentfirst}

 \usepackage[T1]{fontenc}
 \usepackage[unicode=true,pdfusetitle,
  bookmarks=true,bookmarksnumbered=false,bookmarksopen=false,
  breaklinks=false,pdfborder={0 0 1},backref=false,colorlinks=true]
  {hyperref}
 \hypersetup{
  citecolor=blue}
\def\mchi{\rm m_{\chi}}
\def\chiN{\chi {\rm N}}
\def\SIchiN{\rm \chi N ^{SI}}
\def\SRchie{\rm \chi e ^{SR}}
\def\LRchie{\rm \chi e ^{LR}}
\def\chie{\rm \chi e}
\def\SIsigmachiN{\sigma^{\rm SI}_{\chiN}}
\def\LRsigmachie{\sigma^{\rm LR}_{\chie}} 
\def\SRsigmachie{\sigma^{\rm SR}_{\chie}} 
\def\keVee{\rm keV_{ee}}
\def\err3rms{{\pm} 3 \sigma}
\def\dofchi2{ \chi^2 / {\rm dof} }

\def\f3{{\Gamma}_{tot} ^ {( 1 \chi , 2 \chi)}}
\def\1chibf{{\Gamma}_{tot}^{1 \chi}}
\def\2chibf{{\Gamma}_{tot}^{2 \chi}}
\def\fN{{\Gamma}_{\rm N}^0}

\def\C3{{\rm C}_{tot}}
\def\DL{\rm DL}
\def\T{\rm T}

\newcommand{\as}{Institute of Physics, Academia Sinica,
Taipei 11529.}
\newcommand{\bhu}{Department of Physics, Institute of Science,
Banaras Hindu University,
Varanasi 221005.}
\newcommand{\cusb}{Department of Physics,
School of Physical and Chemical Sciences,
Central University of South Bihar, Gaya 824236.}
\newcommand{\ntu}{
Department of Physics, CTP and LeCosPA, National Taiwan University,
Taipei 10617.}
\newcommand{\ndhu}{
Department of Physics, National Dong Hwa University,
Shoufeng, Hualien 97401.}
\newcommand{\glau}{Department of Physics,
Institute of Applied Sciences and Humanities,
GLA University, Mathura 281406. } %% India.}

\newcommand{ \hnbgu }{ Department of Physics, H.N.B. Garhwal University,
Srinagar, Uttarakhand 246174.} %% India. }
\newcommand{\deu}{Department of Physics,
Dokuz Eyl\"{u}l University, Buca, Izmir 35160.} %% Turkey.}
\newcommand{\itu}{Department of Physics Engineering,
Istanbul Technical University, Sariyer, Istanbul 34467.} %% Turkey.}
\newcommand{\thu}{Department of Engineering Physics, Tsinghua University,
Beijing 100084. }
\newcommand{\scu}{College of Physics,
Sichuan University, Chengdu 610065.}

\newcommand{\corrhtw}{htwong@phys.sinica.edu.tw}
\newcommand{\corrlhb}{lihb@gate.sinica.edu.tw}
\newcommand{\corrpandey}{pandey2148@g.ntu.edu.tw}
\newcommand{\corrsingh}{manu@gate.sinica.edu.tw}

\begin{document}

\title{
Dark Matter Annual Modulation Analysis
with Combined Nuclear and Electron Recoil Channels
}

\author{ H.B.~Li } \altaffiliation[Corresponding Author: ]{ \corrlhb } \affiliation{ \as }
\author{ M.K.~Pandey }  \altaffiliation[Corresponding Author: ]{ \corrpandey } \affiliation{ \ntu }
\author{ C.H.~Leung }  \affiliation{ \as }
\author{ L.~Singh }  \affiliation{ \cusb } \affiliation{ \as }
\author{ H.T.~Wong } \altaffiliation[Corresponding Author: ]{ \corrhtw } \affiliation{ \as }
\author{ H.-C.~Chi }  \affiliation{ \ndhu }
\author{ M.~Deniz } \affiliation{ \deu } \affiliation{ \as}
\author{ Greeshma~C. }  \affiliation{ \as } \affiliation{ \cusb }
\author{ J.-W.~Chen }  \affiliation{ \ntu }
\author{ H.C.~Hsu } \affiliation{ \as }
\author{ S.~Karada\v{g} } \affiliation{ \as } \affiliation{ \itu }
\author{ S.~Karmakar }  \affiliation{ \as } \affiliation{ \glau }
\author{ V.~Kumar }  \affiliation{ \as } \affiliation{ \glau }
\author{ J.~Li } \affiliation{ \thu }
\author{ F.K.~Lin } \affiliation{ \as }
\author{ S.T.~Lin } \affiliation{ \scu } \affiliation{ \as }
\author{ C.-P.~Liu }  \affiliation{ \ndhu }
\author{ S.K.~Liu } \affiliation{ \scu }
\author{ H.~Ma }  \affiliation{ \thu }
\author{ D.K.~Mishra }  \affiliation{ \cusb } \affiliation{ \as }
\author{ K.~Saraswat } \affiliation{ \as }
\author{ V.~Sharma }  \affiliation{ \hnbgu } \affiliation{ \as }
\author{ M.K.~Singh } \altaffiliation[Corresponding Author: ]{ \corrsingh } \affiliation{ \as } \affiliation{ \bhu }
\author{ M.K.~Singh } \affiliation{ \glau } \affiliation{ \as} 
\author{ V.~Singh }  \affiliation{ \cusb } \affiliation{ \bhu } \affiliation{ \as }
\author{ D.~Tanabe } \affiliation{ \as }
\author{ J.S.~Wang } \affiliation{ \as }
\author{ C.-P.~Wu }  \affiliation{ \ntu } \affiliation{ \ndhu }
\author{ L.T.~Yang } \affiliation{ \thu }
\author{ C.H.~Yeh } \affiliation{ \as }
\author{ Q.~Yue } \affiliation{ \thu }

\collaboration{ TEXONO Collaboration }

%% ================================

\date{\today}% It is always \today, today,

\begin{abstract}

After decades of experimental efforts,
the DAMA/LIBRA($\DL$) annual modulation (AM) analysis on the
$\chiN$ (WIMP Dark Matter interactions on nucleus) channel remains
the only one which can be interpreted as positive signatures. 
This has been refuted by numerous time-integrated (TI) 
and AM analysis. 
It has been shown that $\chie$ (WIMP interactions with electrons) alone
is not compatible with the DL AM data.
We expand the investigations by performing
an AM analysis with the addition of $\chie$ 
long-range and short-range interactions to $\chiN$,
derived using the Frozen Core Approximation method.
Two scenarios are considered, where the $\chiN$ and $\chie$ processes
are due to a single $\chi$ ($\1chibf$) or two different $\chi$'s ($\2chibf$).
The combined fits with $\chiN$ and $\chie$ provide 
stronger significance to the $\DL$ AM data which are compatible with the
presence of additional physical effects beyond $\chiN$ alone.
This is the first analysis which explores how $\chie$ AM can play
a role in DL AM.
The revised allowed regions as well as
the exclusion contours from the other null AM experiments 
are presented. 
All DL AM allowed parameter spaces in  $\chiN$ and $\chie$ channels
under both $\1chibf$ and $\2chibf$
are excluded at the 90\% confidence level by the combined null AM results.
It can be projected that DL-allowed parameter spaces from 
generic models with interactions induced by two-WIMPs
are ruled out. 
\end{abstract}

%% PACS, the Physics and Astronomy % Classification Scheme.
%%\pacs{
%%95.35.+d,
%%}
%Use showkeys class option if keyword %display desired
%%\keywords{
%%Dark matter,
%%}

\maketitle

%% =======  Introduction ===========

\section{Introduction}

There are compelling experimental evidence that about 
one-quarter of the energy density of the Universe 
is composed of dark matter (DM),
whose exact nature and properties remain unknown. 
Searches of DM in numerous
directions with diverse techniques 
are intense areas of fundamental research~\cite{PDG2020}.
A favored DM candidate is the 
weakly interacting massive particle 
(WIMP, denoted as $\chi$)~\cite{NILLES19841, pnas.1308787111}.

Direct experimental searches assume finite interactions between 
WIMPs with electrons ($\chie$) and nuclei ($\chiN$).
Positive signatures manifest as 
excess events over known background in the measured ``time-integrated'' (TI) energy spectra 
and in their annual modulation (AM) due to
changes of the relative velocity between the Earth and the WIMPs in 
the galactic halo~\cite{Drukier1986}. 
TI analysis is sensitive to uncertainties of background modeling
while AM analysis only requires the background is stable with time
but independent of other details.

%%  ------------------------

After decades of experimental efforts, 
the only result consistent with positive WIMP signatures 
is from the AM analysis on $\chiN$ 
from the DAMA/LIBRA($\DL$) experiment with NaI(Tl) 
scintillating crystals~\cite{BERNABEI2008297}. 
This interpretation, however, is challenged and rejected by 
numerous $\chiN$ experiments with TI analysis~\cite{PDG2020}
and several with 
AM analysis~\cite{xmass2018,cdexam2019,ANAIS112_2021,cosine2024,anais2023}
using a variety of targets.  
In particular, the COSINE~\cite{cosine2024} and 
ANAIS~\cite{anais2023} experiments
adopt the same NaI(Tl) as target nuclei as DL.
There are attempts to explain the $\DL$ AM data
with scenarios other than $\chiN$ detection,
such as complications in the 
analysis procedures~\cite{cosine100_2018,Buttazzo2020}.

%% =====  Table 1 :  AM Experiments ==
%% CDEX         1107.5 kg.d
%% ANAIS        322.82 kg.y
%% COSINE-100   358 kg.yr

\begin{table*}[!htbp]
 \centering
 \caption{
The AM experiments whose data are selected for this analysis.
Their relevant experimental configurations are listed.
}
\begin{center}
\renewcommand{\arraystretch}{1.1}
%% \begin{tabular}{l|c|c|c|c|c|c|c|c}
\begin{tabular}{lcccccccc}
\hline 
Experiment    & Data Set     & Detector                     
& ~~~~ Mass ~~~~     & Live Time/Duration &  ~~ {Exposure} ~~   & {Threshold}    & Ref.         \\  
& &  & ~~ (kg) ~~   & (year)   & {(ton-year)} & {(keV$_{\rm{ee}}$)} &   \\  \hline \hline
DAMA/LIBRA & DAMA   & $\multirow{3}{*}{NaI (Tl)}$  
& 87.3    & 3.32/6.85    & 0.29  & 2.0    &     \\
~~~(DL)~~~ & DL-Phase~1 &   
& 242.5      & 4.29/7.10    & 1.04      & 2.0    & \cite{dama2019}     \\
& DL-Phase~2&   & 242.5  & 4.66/5.89   & 1.13 & 1.0  &   \\ \hline 
COSINE & ~~ COSINE-100  ~~  & NaI (Tl)                     
& 61.3    & 5.80/6.40    & {0.358}    & {0.7}   & \cite{cosine2024}   \\
ANAIS  & ANAIS-112 & NaI (Tl)                     
& 112.5  & 2.87/3.03  & {0.323}   & {1.0}   & \cite{CoarasaCasas:2021euy,anais2023}   \\ 
CDEX  & CDEX-1B\ & Ge  
& 0.939      & 3.20/4.20    & {0.00303}   & {0.25}   & \cite{cdexam2019}   \\
XMASS & XMASS-I &  Liquid Xe   
& 832.0      & 2.19/2.70          & {1.82}                  & {1.0}             & \cite{xmass2018}    \\ \hline 
\end{tabular}
\end{center}
\label{tab::am_expt}
\end{table*}

%% ===========

As elaborated in Section~\ref{sect::results} and summarized in
Table~\ref{tab::DAMA}, 
there exists tension to interpret the DL data at ($1 {-} 4 ~ \keVee$)
exclusively by a $\chiN$-only scenario.
The p-value of 0.07 implies additional effects
are preferred that contribute to the DL AM low energy data.

Analogously, it has been demonstrated~\cite{PhysRevD.100.063017}
$-$ and verified in our studies $-$ that
AM analysis with the $\chie$-only channel
%% (setting $\SIsigmachiN {=} 0$ in Eq.~\ref{eq::3d_fit})
is unable to provide an acceptable fit to
the $\DL$ data.
The AM rates are negligible above $3 ~ \keVee$ in 
relevant range for this work of $\mchi {<} 10^3 ~ {\rm GeV}$, 
in serious discrepancy with the data. 

In comparison, when a $\chiN {+} \chie$ scenario
is considered in AM analysis,
a p-value of 0.008 is obtained
for its $\chi2$-difference with the $\chiN$-only case.
This indicates that the addition of  $\chie$ interactions to
$\chiN$ is statistically significant to match the DL low energy AM data. 

This article is organized as follows.
We discuss the selection of input data in this analysis
in Section~\ref{sect::formulation}, followed by 
the evaluation of the $\chie$ cross sections and
the statistical analysis procedures. 
The results and their interpretations are presented
in Section~\ref{sect::results}, in which
the two cases of the $\chiN$ and $\chie$ interactions 
originated from the same or different $\chi$s
are treated separately.

%% ========= Figure 1 =======

\begin{figure}%%[hbt]
\begin{center}
{\bf (a)}\\
\includegraphics[width=0.85\linewidth]{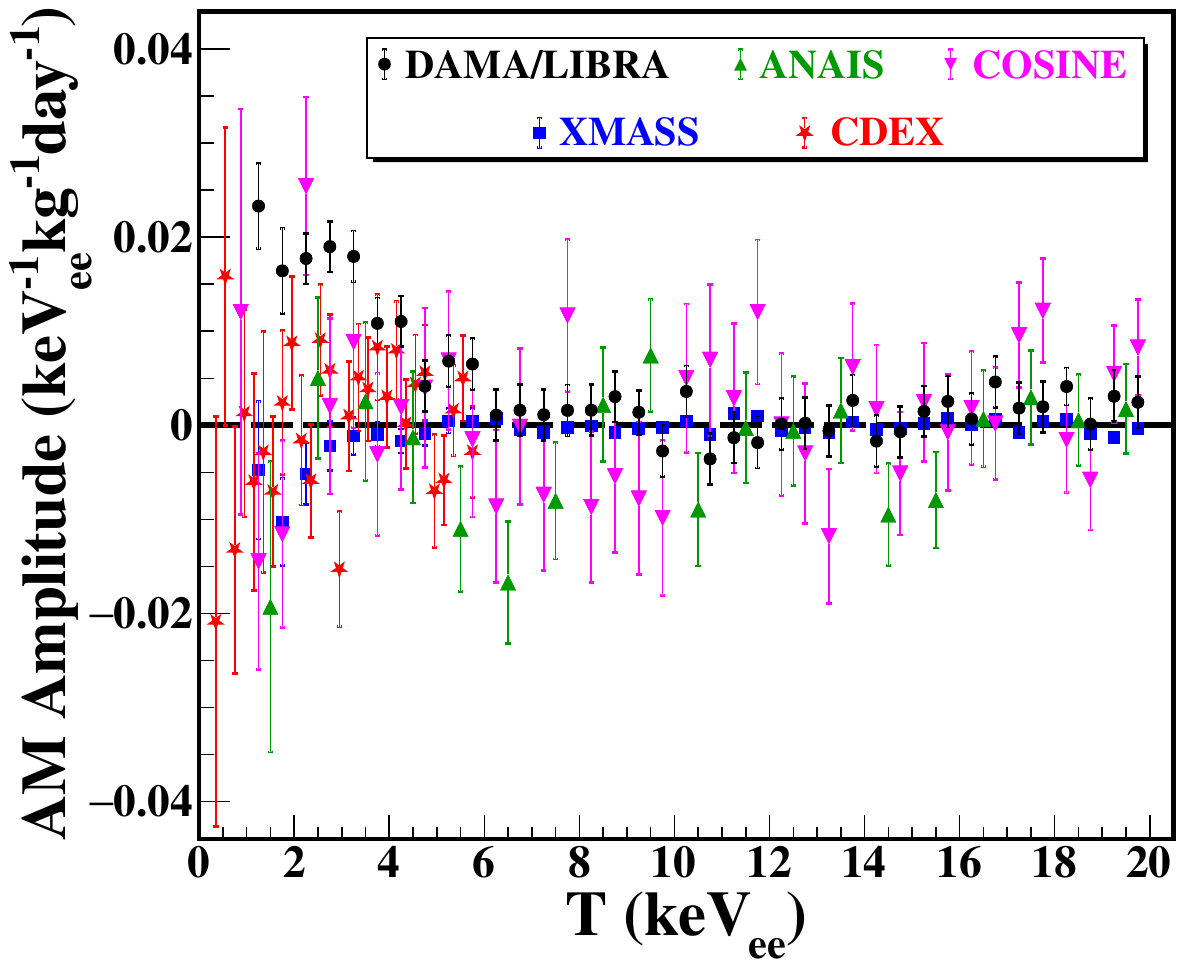}\\
{\bf (b)}\\
\includegraphics[width=0.85\linewidth]{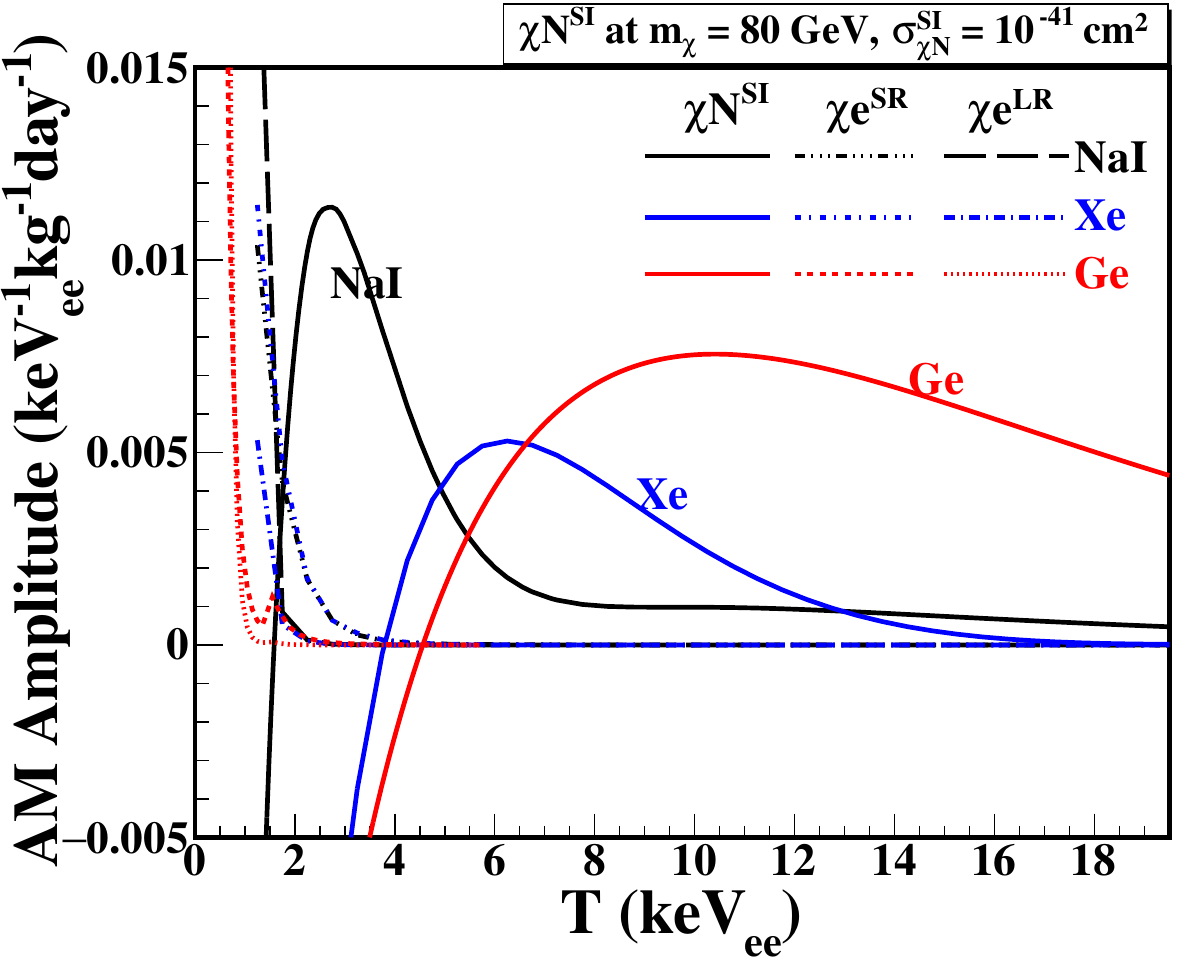}\\
{\bf (c)}\\
\includegraphics[width=0.85\linewidth]{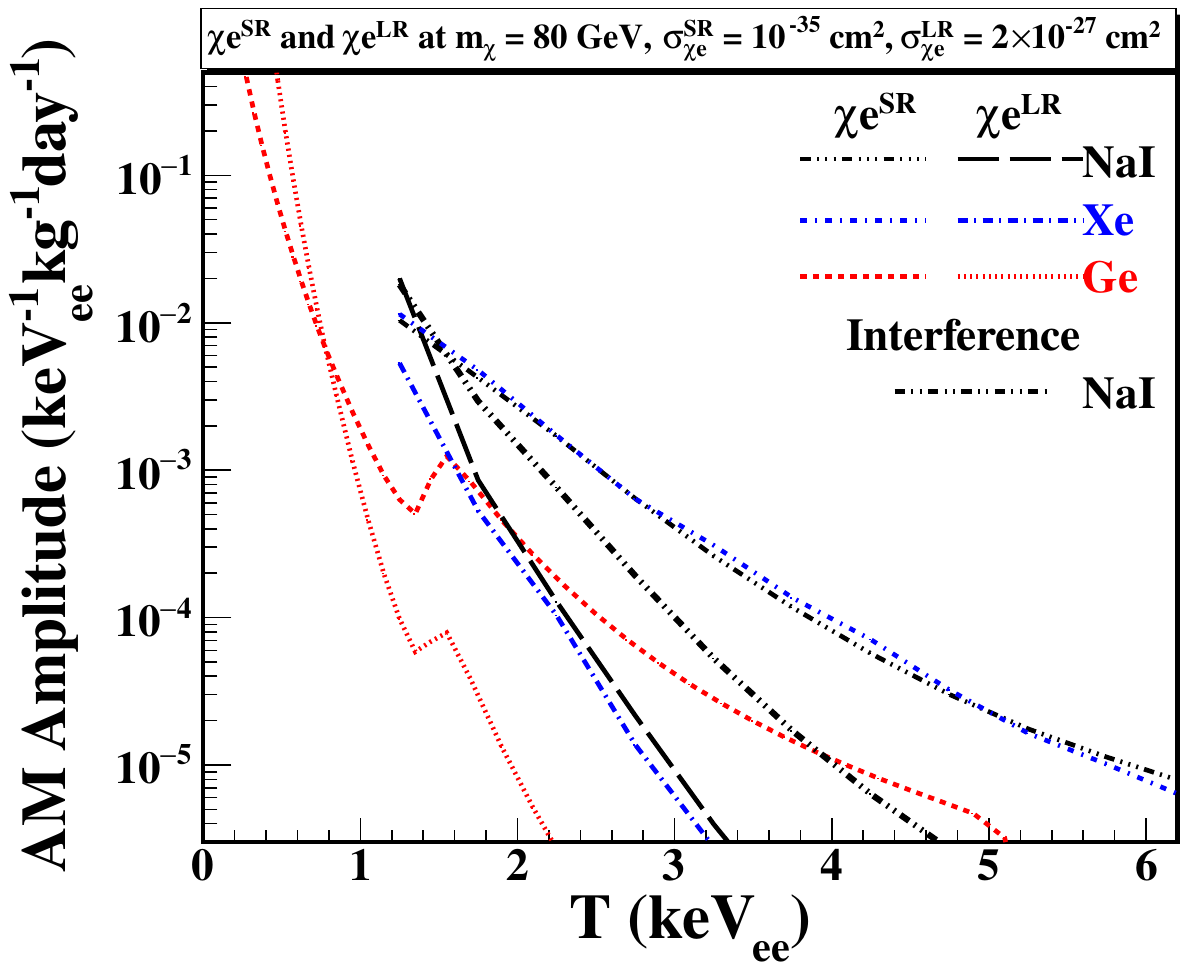}
\end{center}
\caption{
(a)
Published AM amplitudes data from DM experiments 
with NaI in
$\DL$~\cite{dama2019},
COSINE~\cite{cosine2024} and
ANAIS~\cite{CoarasaCasas:2021euy,anais2023},
Xe in XMASS~\cite{xmass2018} and
Ge in CDEX~\cite{cdexam2019}.
To allow the different nuclei
be presented with a common axis for display purposes,
the depicted CDEX event rates 
are re-scaled by a factor 0.04.
%% to account for the differences in bin-sizes and target mass number squared. 
The expected differential AM spectra for positive WIMP signatures in:
(b) $\SIchiN$ and (c) in $\LRchie$ and $\SRchie$ 
for all three targets taking the case of 
$\mchi {=} 80 ~ {\rm GeV}$ as example,
together with the interference spectrum 
between $\LRchie$ and $\SRchie$ for NaI.
The $\LRchie$ and $\SRchie$ spectra in (c)
are superimposed as dotted lines in (b)
showing their responses to AM are very different from $\SIchiN$.
}
\label{fig::data}
\end{figure}

%% ========= Figure 2 ==========

\begin{figure}%%[hbt]
\begin{center}
{\bf (a)}\\
\includegraphics[width=0.80\linewidth]{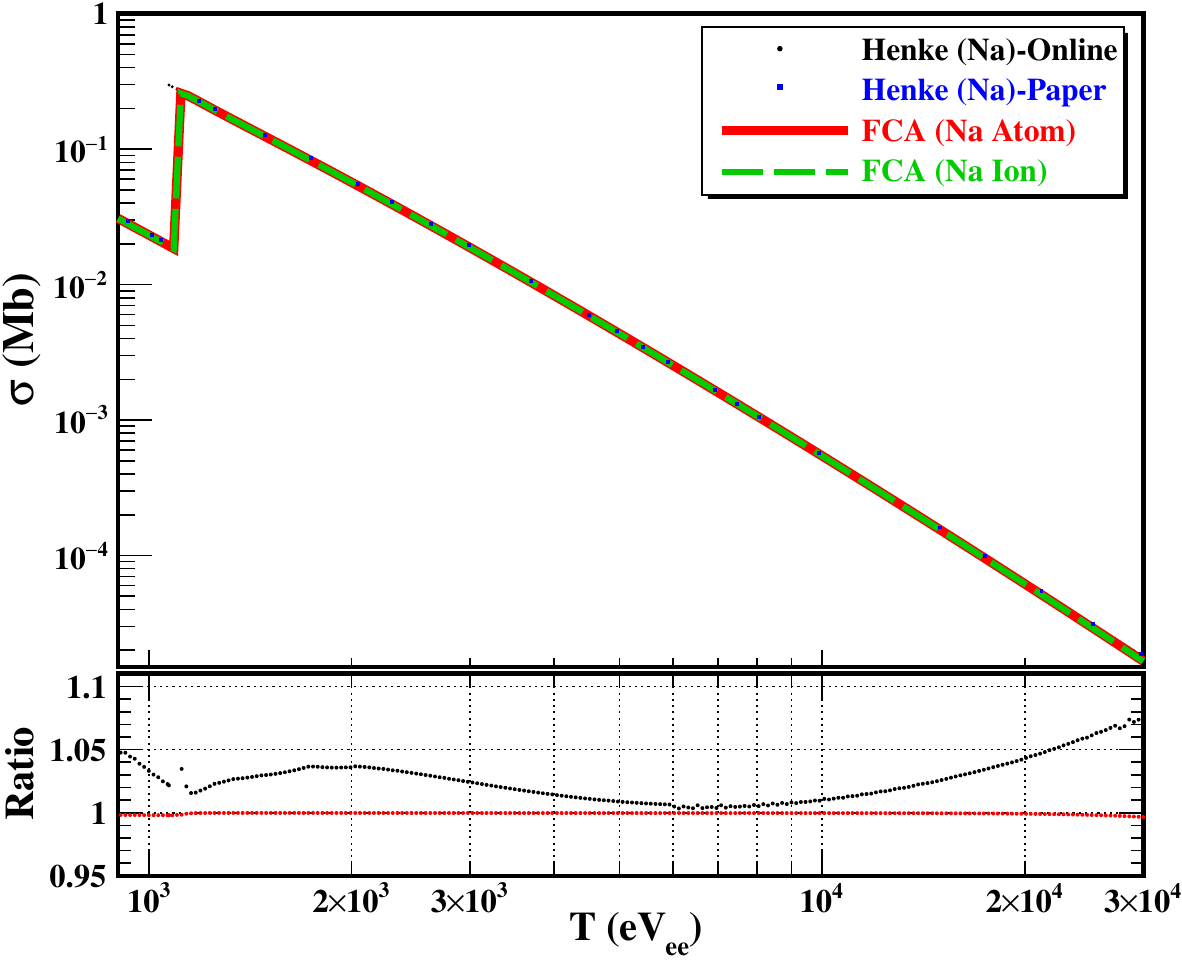}\\
{\bf (b)}\\
\includegraphics[width=0.80\linewidth]{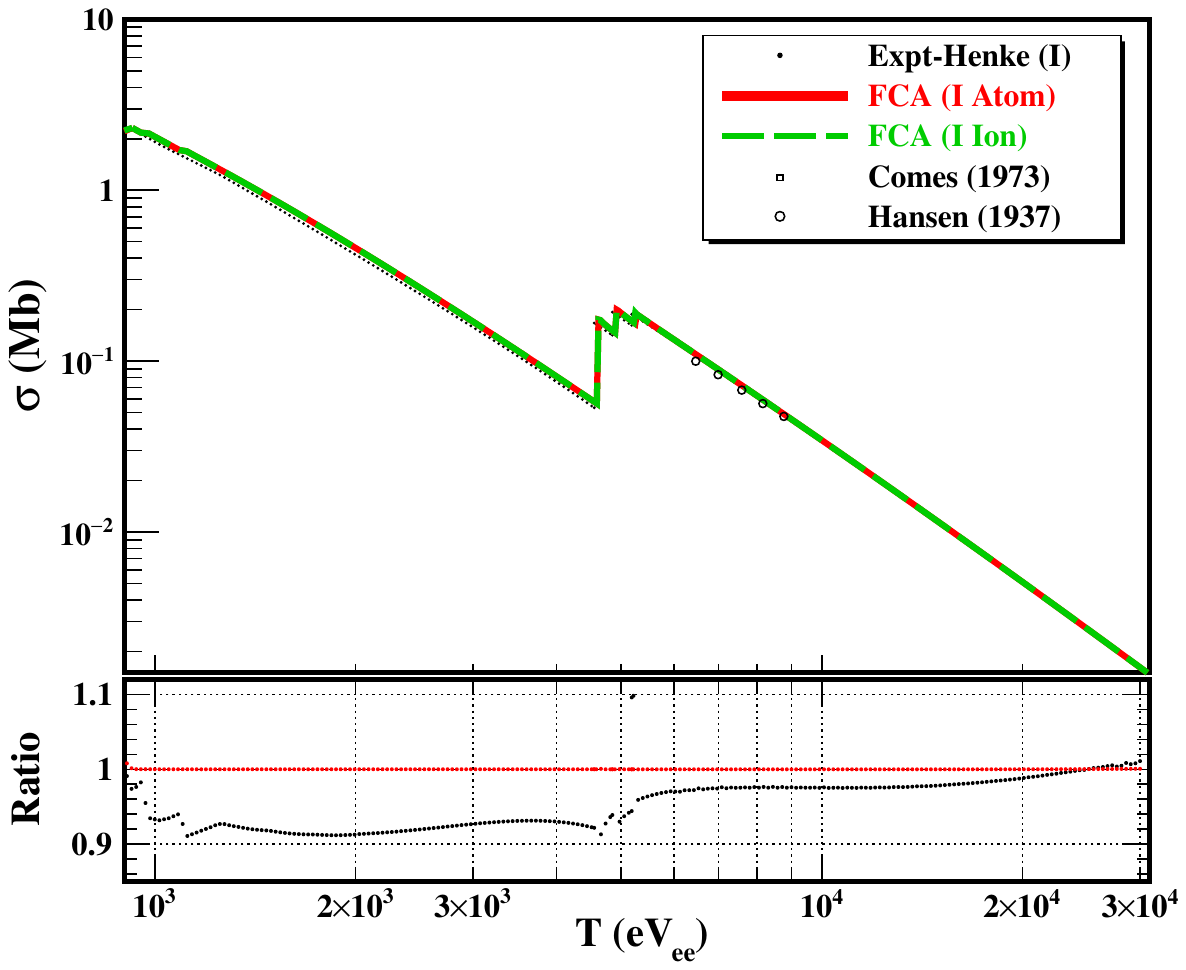}
\end{center}
\caption{
Photo-ionization cross-sections derived by
FCA following Ref.~\cite{pandey2020}
for (a) Na and (b) I,
both as atoms and ions, within
1~$\keVee$ to 30~$\keVee$ energy transfer. 
The experimental data~\cite{Henke:1993eda}
and theoretical values~\cite{hansen1937,comes1973} are superimposed.
Relative deviations from benchmark data are displayed.
}
\label{fig::FCA_NaI}
\end{figure}

%% ===========  Formulation ==============

\section{Formulation}
\label{sect::formulation}

We performed an analysis 
including {\it both} $\chie$ and $\chiN$ interactions.
The combined analysis would provide improved 
statistical fit to the $\DL$ data. 
Three types of WIMPs interactions are included:
spin-independent WIMP-nucleon interaction ($\SIchiN$), 
long and short-range WIMP-electron interaction 
($\LRchie$ and $\SRchie$, respectively)~\cite{pandey2020}.
Their cross-sections (respectively $\SIsigmachiN$, $\LRsigmachie$ and $\SRsigmachie$)
are all functions of $\mchi$. 

\subsection{Input Data}

The input to the analysis, 
as depicted in Figure~\ref{fig::data}a,
are the published AM amplitudes data. 
A summary of the relevant features of the AM experiments 
is given in Table~\ref{tab::am_expt}.
These AM data sets are complementary in their strength in probing different parameter space. 
The $\DL$~\cite{dama2019} experiment is the only one reporting positive AM results.
COSINE~\cite{cosine2024} and ANAIS~\cite{CoarasaCasas:2021euy,anais2023} 
use the same detector technology $-$ NaI(Tl) crystal scintillators $-$
with identical target isotopes as $\DL$.
Comparisons would be model-independent in principle,
though the variations on quenching factors among different NaI(Tl) crystals
have been raised~\cite{Ko_2019}.
CDEX is distinguished with low detector threshold ($0.25 ~ \keVee$ for the AM analysis)
made possible by novel p-type point-contact germanium detectors,
while XMASS is a single-phase liquid xenon detector with large exposure (1.82~ton-year).
The electron-equivalent unit $\keVee$ is used
to characterize detector response and 
measured energy ($\T$), unless otherwise stated.

The maximum(minimum) AM amplitudes are on
June 1(December 2)~\cite{dama2008,Froborg_2020}
following the standard DM halo model~\cite{PDG2020,Drukier1986}.
The quenching factors used by $\DL$~\cite{BERNABEI1996757}
is applied to all the three NaI(Tl) experiments.
The expected AM differential spectra for positive WIMP signatures
in $\SIchiN$ for the three targets are displayed in
Figure~\ref{fig::data}b.
The AM spectral shapes  originate from the differences in 
$\chi$-velocity relative to the Earth between summer and winter.
A characteristic feature %%AM spectral shape in $\SIchiN$
is a drop from enhancement to deficit at low energy.
The turning point is $\mchi$-dependent.

%% =================

\subsection{Cross-Sections in $\chie$}

Recent interest in the searches of light DM brings
along intense activities in exploring novel detector concepts. 
Crucial to this development is the $\chie$ detection channels 
which triggers intense research efforts in the 
refinement on the evaluation of its interactions with 
atoms $A$~\cite{pandey2020,PhysRevD.106.063003} in
\begin{equation}
\chi {+} A {\rightarrow} 
\chi {+} e^- {+} A^+  ~~ .
\end{equation}

Frozen Core Approximation (FCA)~\cite{slater_fca, sampson_fca},
is a well-established framework in atomic many-body physics, 
recognized for its effectiveness in unraveling complex interactions and 
dynamics within atomic systems. 
Following our previous work on Ge and Xe 
with FCA~\cite{pandey2020,PhysRevD.106.063003},
we extend the same approach  %%prescribed in Ref.~\cite{pandey2020} 
to Na and I in this work.
The results are validated by comparing the 
derived photo-ionization cross-sections 
against experimental data~\cite{Henke:1993eda}
and previous calculations~\cite{hansen1937,comes1973}.
The photo-ionization cross-sections of Na and I are
illustrated in Figures~\ref{fig::FCA_NaI}a\&b,
showing consistency of theory calculations with measured data to within 5\%
across the relevant energy transfer range of $1 {-} 30 ~ \keVee$.
Identical results are obtained
whether the targets are treated as atoms or ions.
The consistency indicates FCA can provide reliable modeling to the 
interactions of $\chi$ with the atoms.

The differential AM spectra in $\LRchie$ and $\SRchie$ 
for the three targets
are shown in Figure~\ref{fig::data}c.
The $\chie$ recoil energy is shifted lower in winter,
giving rise to rapidly rising AM spectra.
In practice, only data ${<} 4 ~ \keVee$ would contribute
to the analysis of the $\chie$ channels.
The long-range $\chi {\rm e}$ interactions has  
an additional $1/q^{2}$ term where $q$ is 
the 3-momentum transfer~\cite{pandey2020}.
The differential AM spectra therefore rise 
steeper at low recoil energy such that 
studies of $\LRchie$ favor experiments with lower detection threshold. 

The AM spectra in $\LRchie$ and $\SRchie$ for the three targets
are superimposed to the $\SIchiN$ spectra in 
Figure~\ref{fig::data}b, showing their vastly different response
at low energy.

%% ======  Figure 3 ==========

\begin{figure}%%[hbt]
\begin{center}
{\bf (a)}\\
\includegraphics[width=0.85\linewidth]{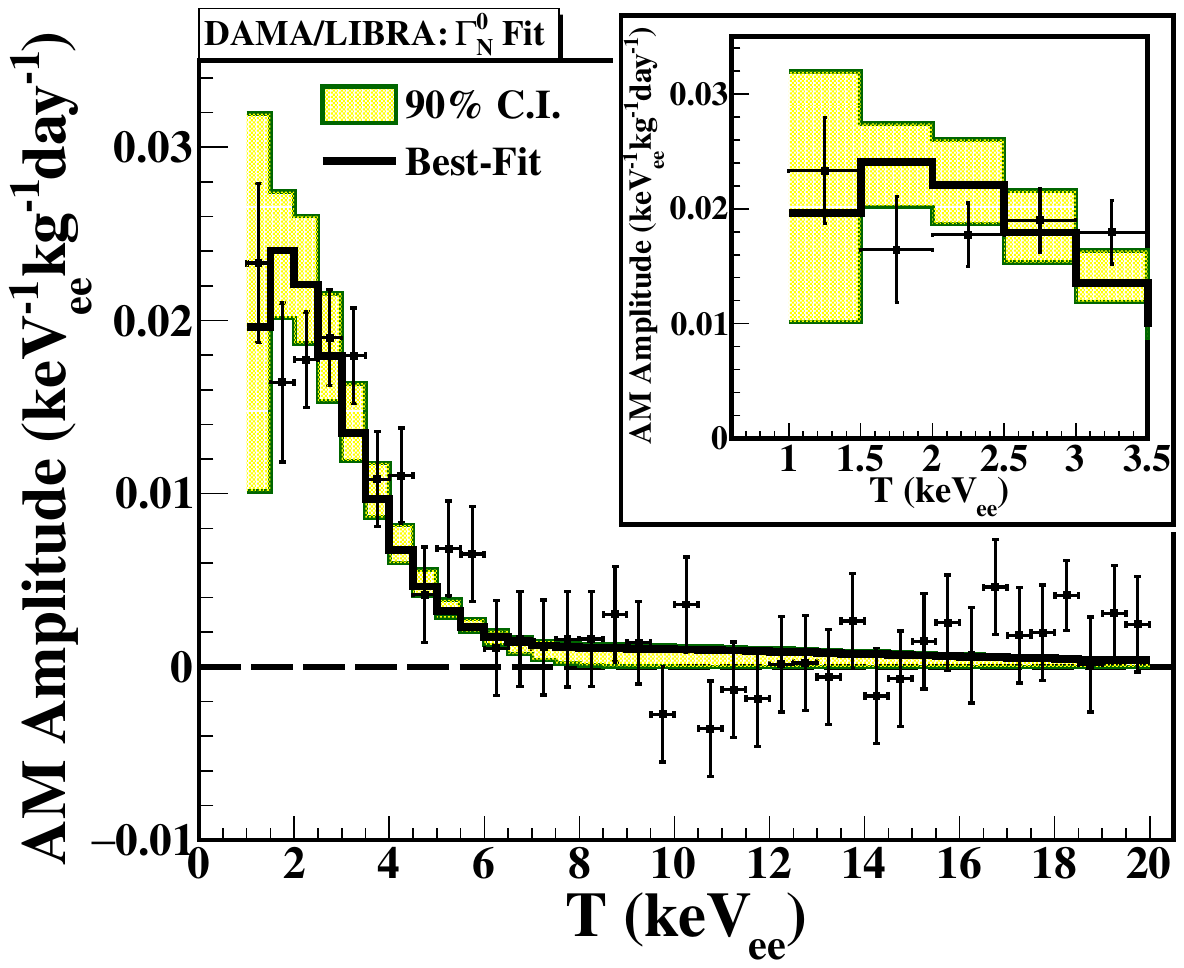}\\
{\bf (b)}\\
\includegraphics[width=0.85\linewidth]{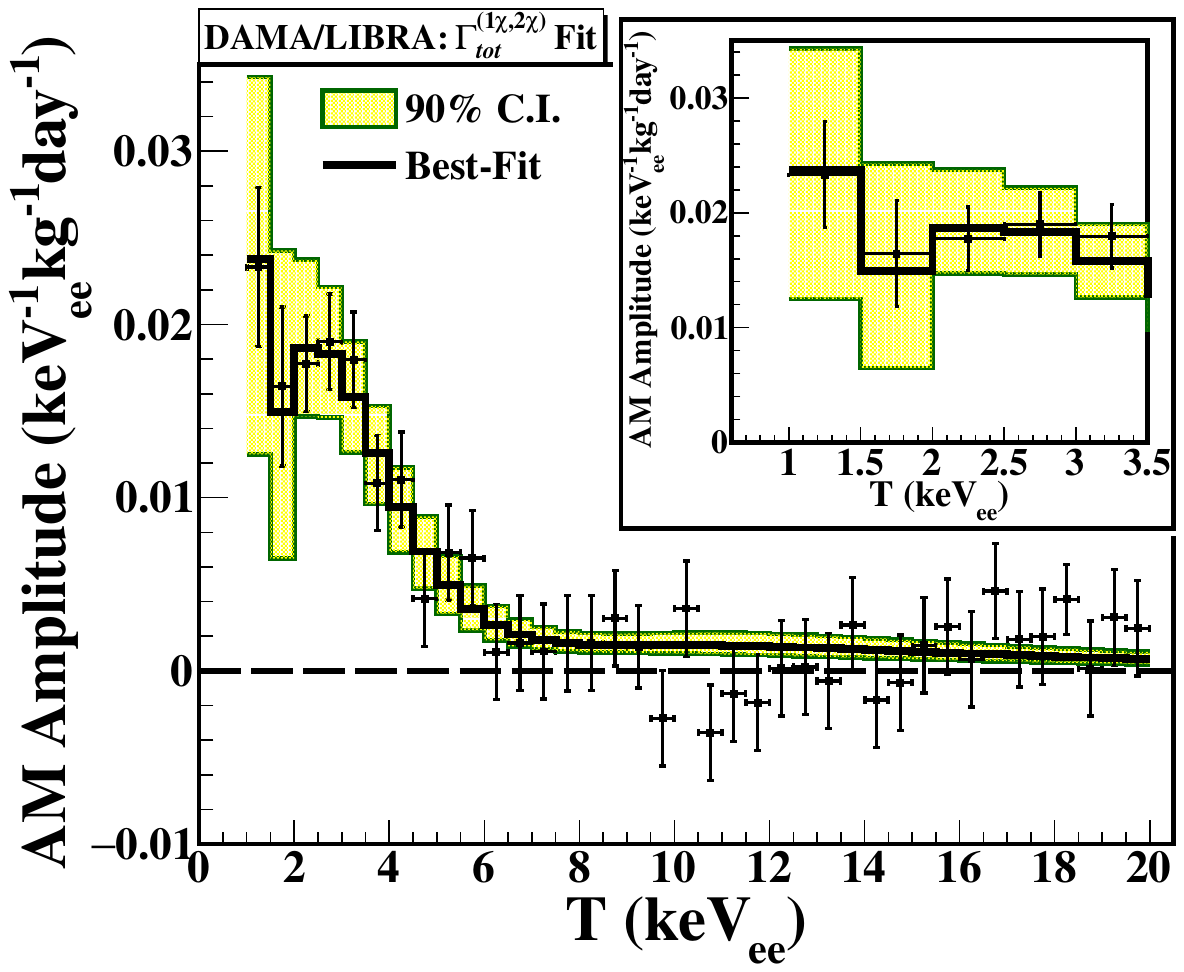}
\end{center}
\caption{
$\DL$ data~\cite{dama2019}
with best-fits under
(a) $\fN$: with $\chiN$ only
where $\mchi {=} 54 ~ {\rm GeV}$,
and
(b) $\f3$:
with all three channels
($\SIchiN {,} \LRchie {,} \SRchie$)
combined, in which case
$\mchi {=} 83 ~ {\rm GeV}$.
}
\label{fig::spectra}
\end{figure}

%% =================

%% =========  Derivation ============

\subsection{Analysis}

The best-fit estimators of the various cross-sections
are derived by a minimum-$\chi ^2$ analysis.
At a given $\mchi$,
\begin{eqnarray}
\label{eq::3d_fit}
\chi ^2 &  {=}  
{\displaystyle \sum_{i}    ~  \frac{1}{\Delta_{i}^{2}} } &
\Big\{ n_{i} -    
\Big[  
{\SIsigmachiN} ~ \phi_{\chi N}^{\rm SI}(\T_{i}) +  
{\LRsigmachie} ~ \phi_{\rm \chi e}^{\rm LR} (\T_{i}) \\   
& & 
+ {\SRsigmachie} ~ \phi_{\rm \chi e}^{\rm SR} (\T_{i}) 
+ {2\sqrt{\LRsigmachie\SRsigmachie}} ~ \phi_{\rm \chi e}^{\rm int} (\T_{i}) 
\Big] \Big\} ^{2}  \nonumber
\end{eqnarray}
where $n_i$ and $\Delta_i$
are the AM amplitudes and uncertainties 
at the $i^{th}$-bin of average energy $\T_{i}$, 
while $\phi_{\rm \chi e}^{\rm SI}$, $\phi_{\rm \chi e}^{\rm LR}$, $\phi_{\rm \chi e}^{\rm SR}$ 
are the normalized spectral functions for
the three interactions. 
The interference spectrum ($\phi_{\rm \chi e}^{\rm int}$) 
between $\LRchie$ and $\SRchie$ 
follows the many-body $\chi$-atom calculations 
given in Eq.~3 of Ref.~\cite{pandey2020},
also depicted in Figure~\ref{fig::data}c.

%% =========================

The $\DL$  data shows positive AM signatures
and reject the null hypothesis with large statistical significance.
Its  best-fit spectra with $\chiN$-channel only 
($\fN {:} ~ {\rm setting} ~
\LRsigmachie {=} \SRsigmachie {=} 0$ in Eq.~\ref{eq::3d_fit}) 
is displayed in Figure~\ref{fig::spectra}a.
This analysis expands to have
all three channels %%($\SIsigmachiN$, $\LRsigmachie$ and $\SRsigmachie$)
taken as the free fitting variables.
Two DM scenarios are considered, parametrized 
by $f_{\chi}$ as the DM relic density fraction
from the $\chi$ interacting via $\chie$:
\begin{enumerate}
\item $\1chibf -$
Both $\chiN$ and $\chie$ interactions are 
due to a single $\chi$ ($f_{\chi} {=} 1$),
such that same constraints on $\mchi$ apply to all channels;
\item $\2chibf -$
The case of independent constraints where
two different $\chi$'s with fractional density 
$f_{\chi}$ and $( 1 {-} f_{\chi} )$
interact separately via $\chie$ and $\chiN$, respectively.
\end{enumerate}
The limiting case of $f_{\chi} {=} 0$ corresponds to the baseline $\fN$.
%% The analysis of $\2chibf$ is applicable to generic interactions 
%% by two independent WIMPs.

%% =======  Table 2 =============

\begin{table}%%[h!]
\begin{center}
\caption{
Comparison of the statistical significance
in terms of $\dofchi2$ of the $\fN$ and $\f3$
fitted to the published DL AM data~\cite{dama2019}.
The p-values which qualify likelihood of tested hypothesis
are also shown.
The results can be compared to those of 
a test case with 20\% reduced uncertainties which
produces p=0.5 on the complete data set (*).
The increased significance from $\fN$ to $\f3$
implies the presence of genuine physical processes.
}
\begin{tabular}{ |c||c|c|c| }
 \hline
 & 
$\fN$: &
$\f3$: &
\multirow{2}{*}{Compare} \\
&
$\SIchiN$ & 
$\SIchiN {+} \LRchie {+} \SRchie$ & \\ \hline
Data &
\multicolumn{2}{c|}{$\dofchi2$} &
$\Delta \dofchi2$  \\ 
$~~ ( \keVee ) ~~$ &
\multicolumn{2}{c|}{(p-value)} &
~~ (p-value) ~~ \\ \hline
\multicolumn{4}{c}{\bf Published Data} \\ \hline
\multirow{2}{*}{1 - 20} &
~~ 32.06/36 ~~ &
22.40/34 &
9.66/2 \\ 
&
(0.66) &
(0.94) &
(0.008) \\ \hline
\multirow{2}{*}{1 - 4} &
8.6/4 &
1.3/2 &
7.26/2 \\ 
&
(0.07) &
(0.52) &
(0.02) \\ 
 \hline
\multicolumn{4}{c}{\bf Test Case (Effects of Reduced Uncertainties)* } \\ \hline
\multirow{2}{*}{1 - 20} &
~~ 48.1/36 ~~ &
33.33/34 &
14.8/2 \\
&
(0.086) &
(0.50)* &
(0.0006) \\ \hline
\multirow{2}{*}{1 - 4} &
12.8/4 &
1.94/2 &
10.86/2 \\
&
(0.012) &
(0.38) &
(0.0044) \\
\hline
\end{tabular}
\label{tab::DAMA}
\end{center}
\end{table}

%% ========================

%% =======  Figure 4 ===========

\begin{figure}%%[hbt]
\begin{center}
{\bf (a)}\\
\includegraphics[width=0.85\linewidth]{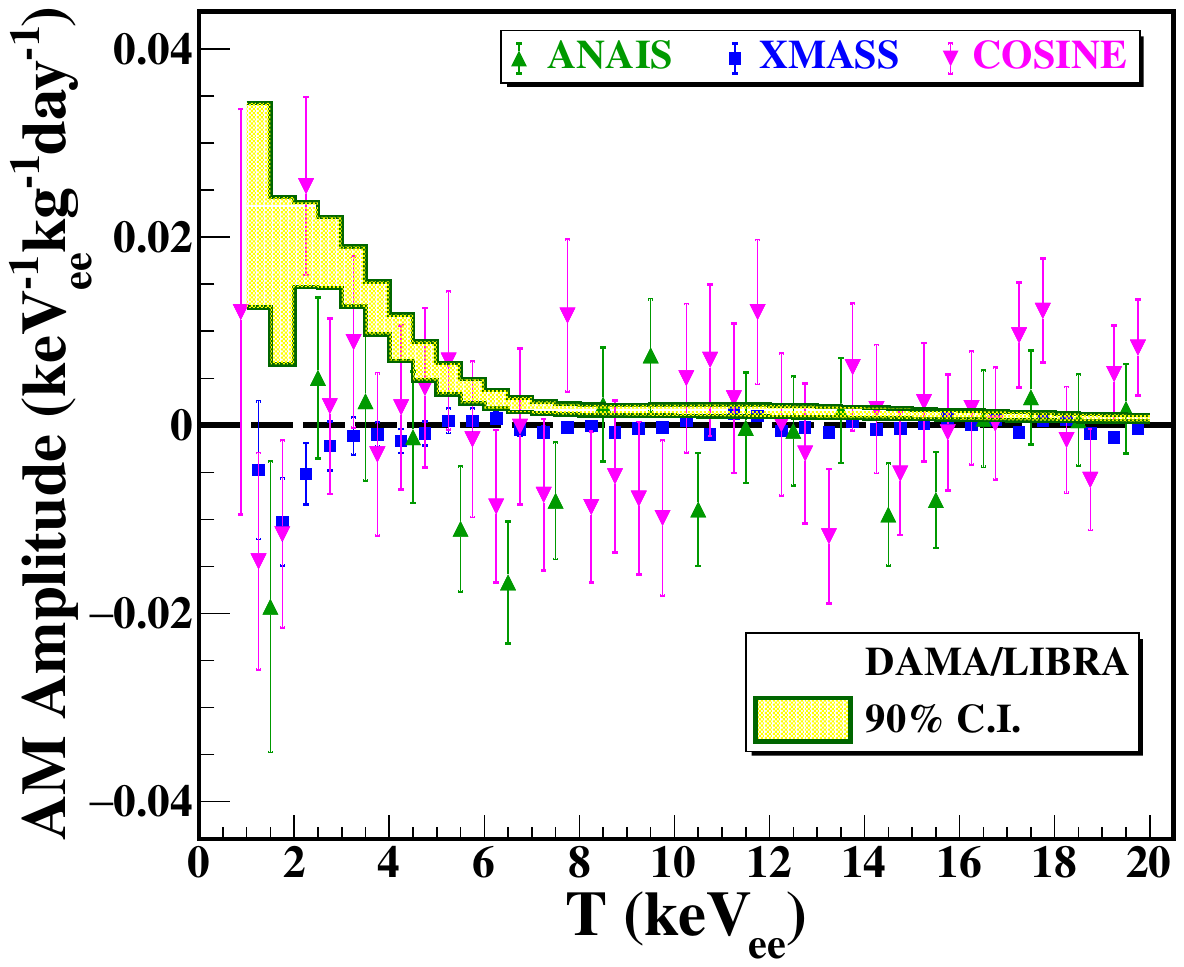}\\
{\bf (b)}\\
\includegraphics[width=0.85\linewidth]{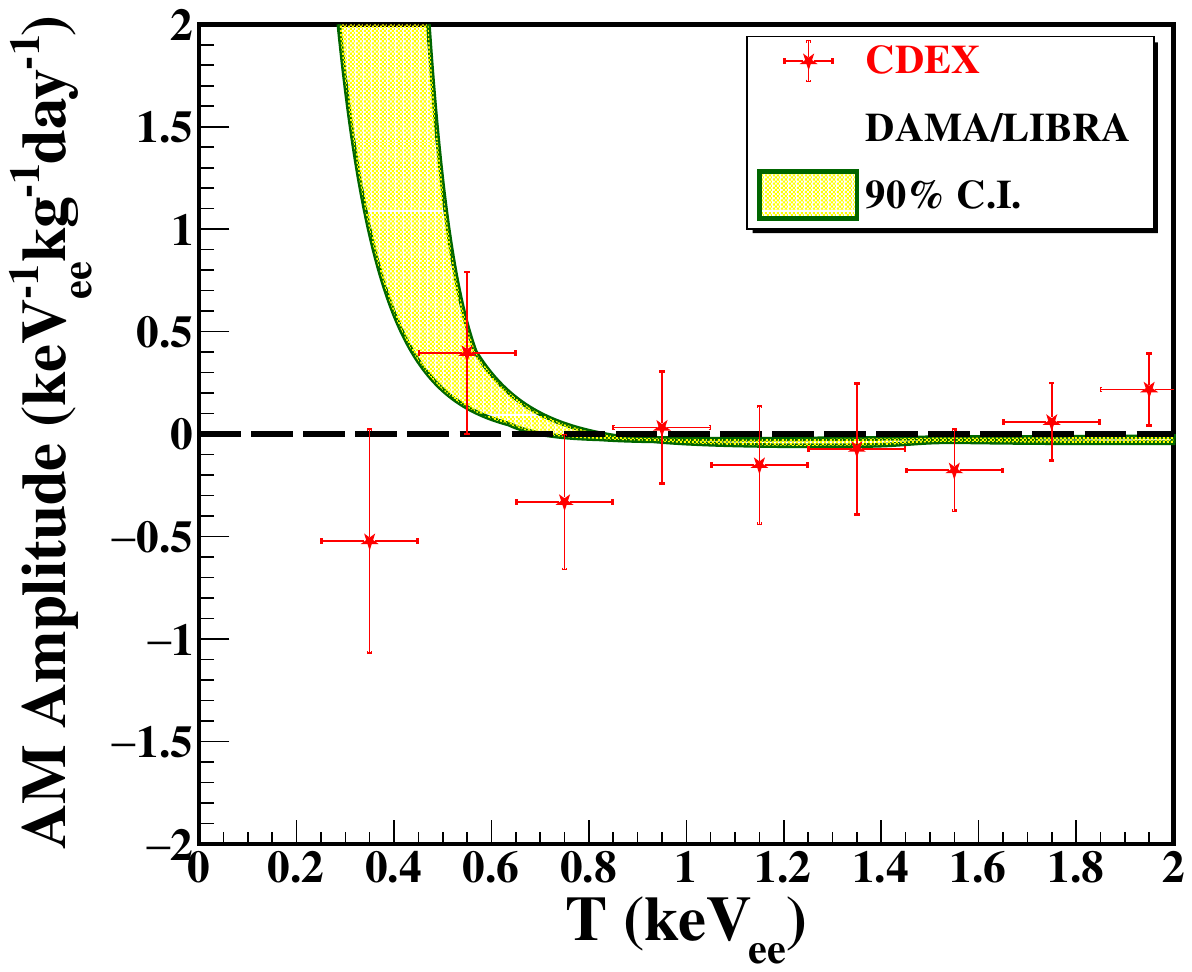}
\end{center}
\caption{
Predicted spectra for the null AM experiments:
(a) COSINE, ANAIS, XMASS, and
(b) CDEX,
due to the $\f3$ best-fit values of ($\SIchiN {,} \LRchie {,} \SRchie$) 
on the DL AM data in Figure~\ref{fig::spectra}.
}
\label{fig::best-fit-projection}
\end{figure}

%% ========== Figure 5 ==========

\begin{figure}%%[hbt]
\begin{center}
{\bf (a)}\\
\includegraphics[width=0.85\linewidth]{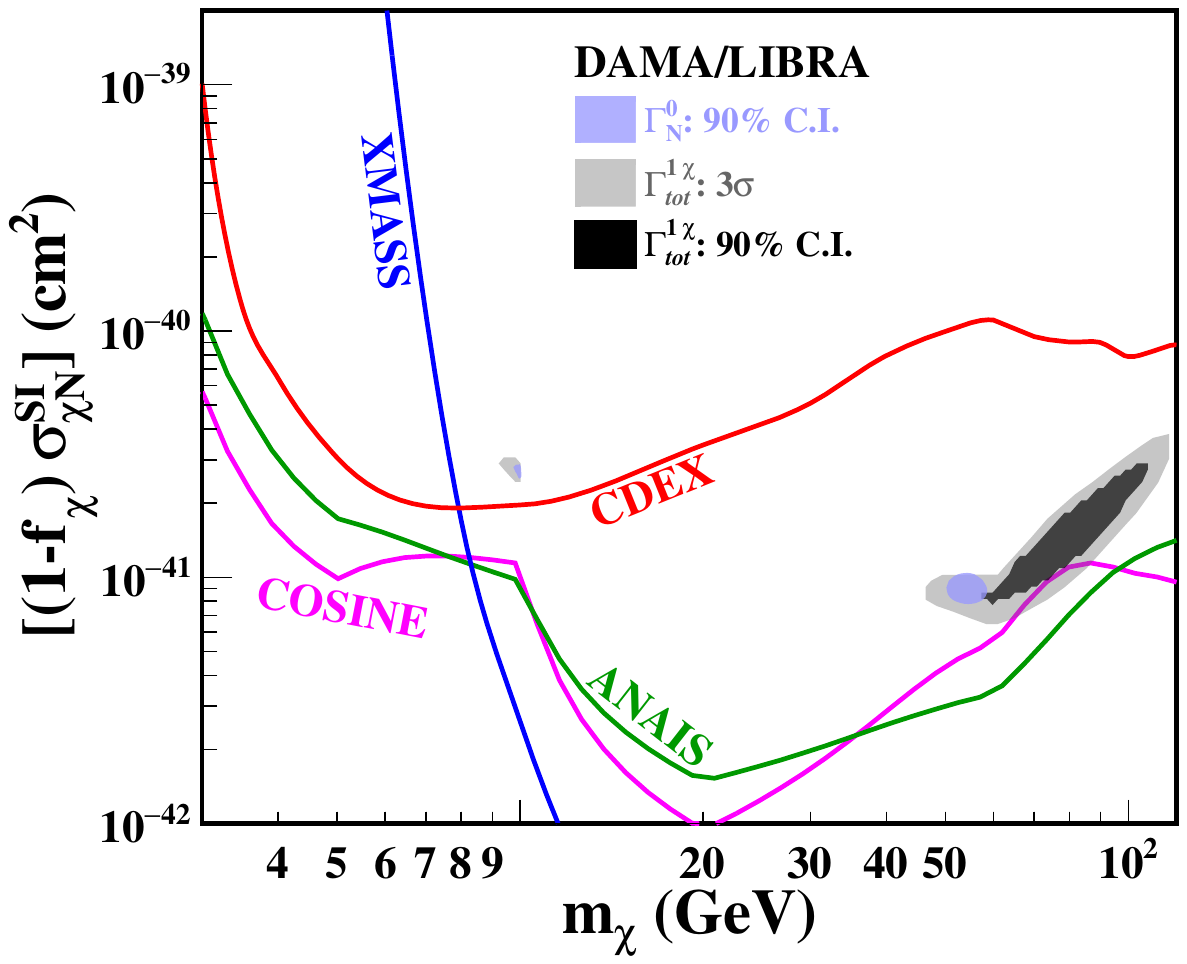}\\
{\bf (b)}\\
\includegraphics[width=0.85\linewidth]{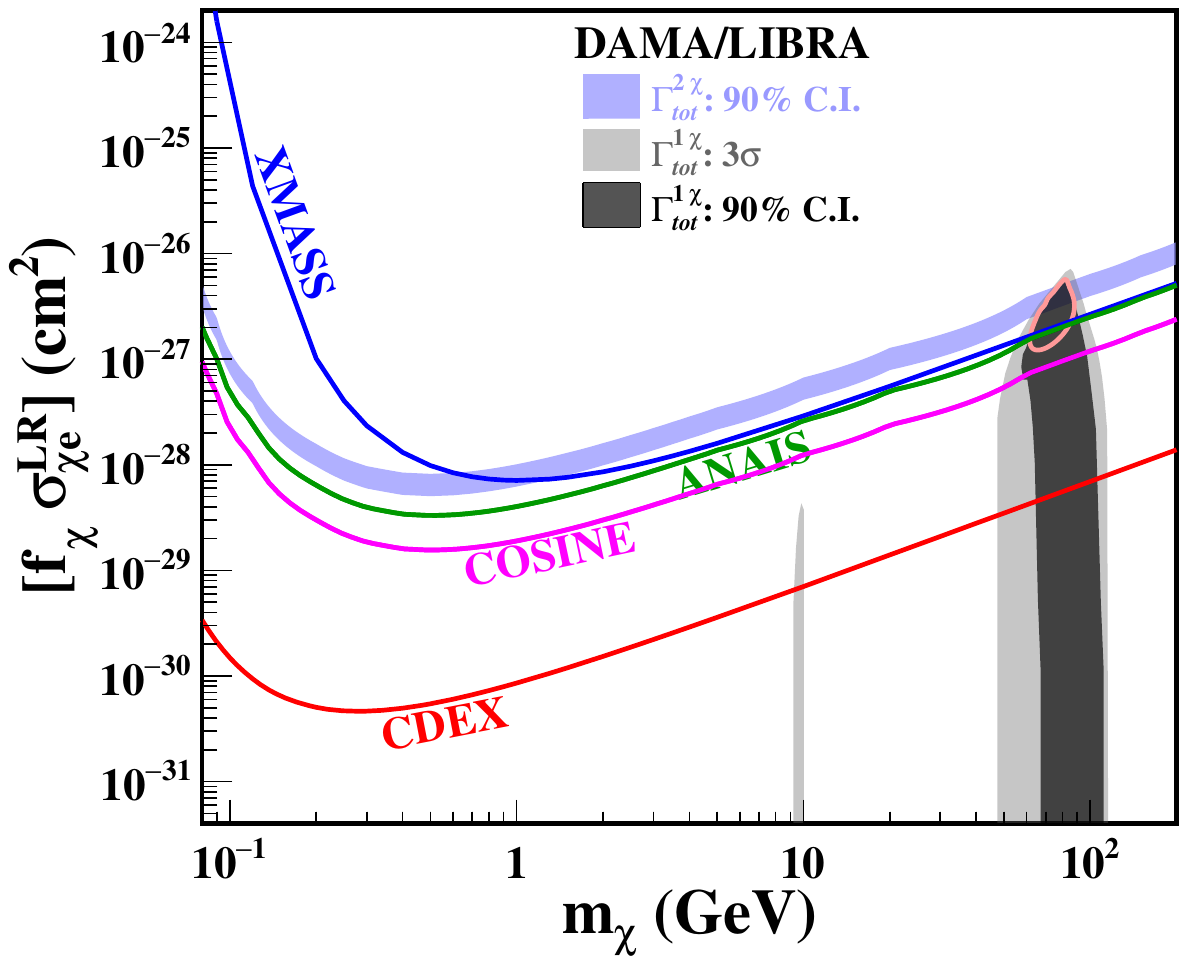}\\
{\bf (c)}\\
\includegraphics[width=0.85\linewidth]{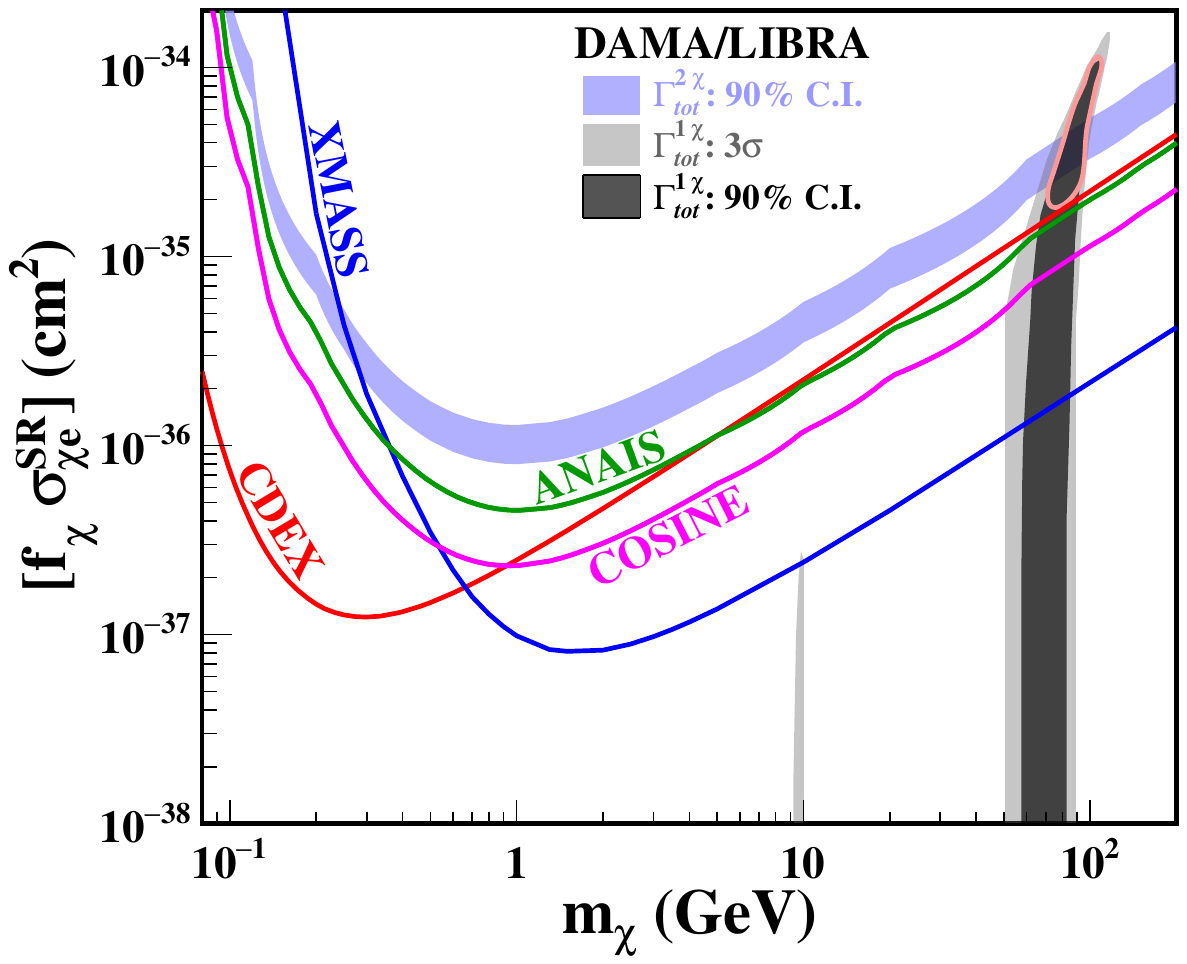}
\end{center}
\caption{
Exclusion plots with AM Data from Na+I, Ge and Xe experiments with
(a) $\SIsigmachiN$
(b) $ \LRsigmachie $, and
(c) $ \SRsigmachie $
versus $\mchi$.
Y-axes are presented in $[f_{\chi} \sigma]$
to accommodate both scenarios $\1chibf$ $( f_{\chi} {=} 1 )$
and $\2chibf$ $( 0 {<} f_{\chi} {<} 1 )$.
The DL allowed regions~\cite{dama2019} from $\1chibf$
at 90\%($\err3rms$) C.I.  are presented as dark(light) shaded areas.
The case of $\fN$ is included in (a) as comparison.
Exclusion contours from the null AM experiments
%% (ANAIS~\cite{CoarasaCasas:2021euy,anais2023}, COSINE~\cite{cosine2024},
%% XMASS~\cite{xmass2018} and CDEX~\cite{cdexam2019})
represent upper limits at 90\% C.L.
The light red circles of (b,c) are allowed regions at 90\% C.I. from $\1chibf$
for ($\LRsigmachie$ at $\SRsigmachie {=} 0$,$\SRsigmachie$ at $\LRsigmachie {=} 0$)
while the light blue bands correspond to those from $\2chibf$.
}
\label{fig::explot}
\end{figure}

%% =========================

%% =========  Results =========

\section{Results and Interpretations}
\label{sect::results}

The combined best-fit in $\1chibf$
is given in Figure~\ref{fig::spectra}b.
The spectra for $\2chibf$ would be identical.
The allowed bands of $\SIsigmachiN$ are derived with
the Wilks' approximation~\cite{Algeri2020}.
The addition of the $\chie$-channels
provides better description of the DL AM data ${<} 4 ~ \keVee$.
The $\f3$-fit is a more inclusive and general form than
the conventional $\fN$.

As illustrated in Table~\ref{tab::DAMA},
the analysis reveal that the interpretation of the
$\DL$ published data~\cite{dama2019} incorporating $\f3$
has higher statistical significance than $\fN$ alone.
The low energy data ($1 {-} 4~\keVee$) gives
p-values of 0.52 for $\f3$
but only 0.07 for $\fN$, indicating that
the $\SIchiN$ channel as the sole physics scenario
cannot explain the low energy data.
In addition, the differences in $\dofchi2$ between $\fN$
and $\f3$
(7.26/2 for ${\rm T} {=} 1 {-} 4~\keVee$
and 9.66/2 for ${\rm T} {=} 1 {-} 20 ~ \keVee$)
correspond to p-values of 0.02 and 0.008, respectively.
This implies the necessity of having
additional physical processes such as 
$\LRchie$ and $\SRchie$  
to explain the AM spectrum~\cite{delta_chi2}.

We note the $\dofchi2$ value for the
complete 1$-$20~keV data set in $\f3$ in
Table~\ref{tab::DAMA} is significantly less than 1. 
This suggests a scenario where the published uncertainties~\cite{dama2019} 
are over-estimated.
We investigate a test case where the uncertainties of the DL data
are uniformly reduced by 20\%, resulting in p=0.5. 
The changes to $\dofchi2$  and p-values 
are shown in Table~\ref{tab::DAMA}. 
The tension against $\fN$ as a valid hypothesis is stronger 
while $\f3$ shows perfect agreement with data.

%% ====================

%% ========  Figure 6 ==========

%% $\DL$~\cite{dama2019},
%% COSINE~\cite{cosine2024} and
%% ANAIS~\cite{CoarasaCasas:2021euy,anais2023},
%% XMASS with Xe~\cite{xmass2018} and
%% CDEX with Ge~\cite{cdexam2019},

\begin{figure}%%[hbt]
\begin{center}
{\bf (a)}\\
\includegraphics[width=0.85\linewidth]{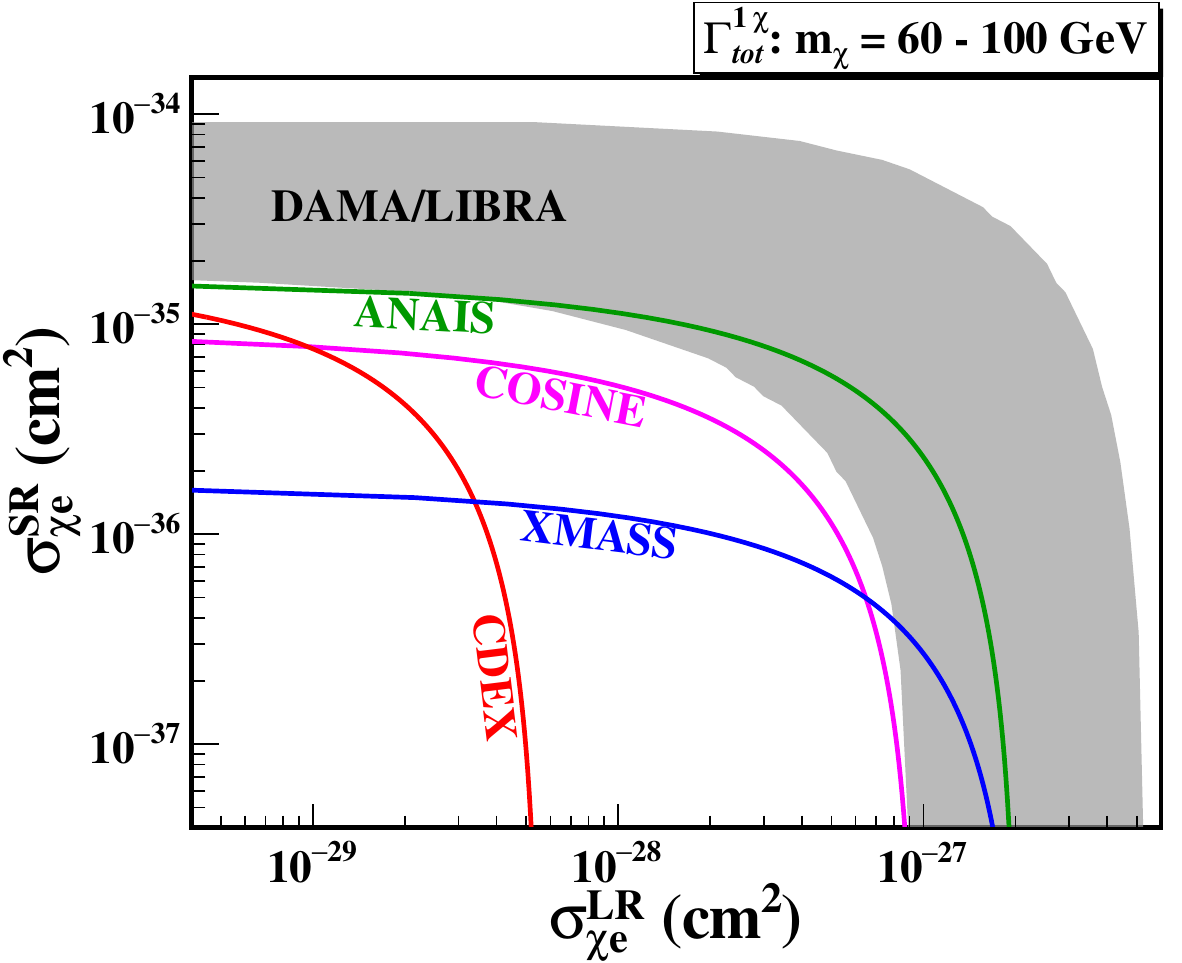}\\
{\bf (b)}\\
\includegraphics[width=0.85\linewidth]{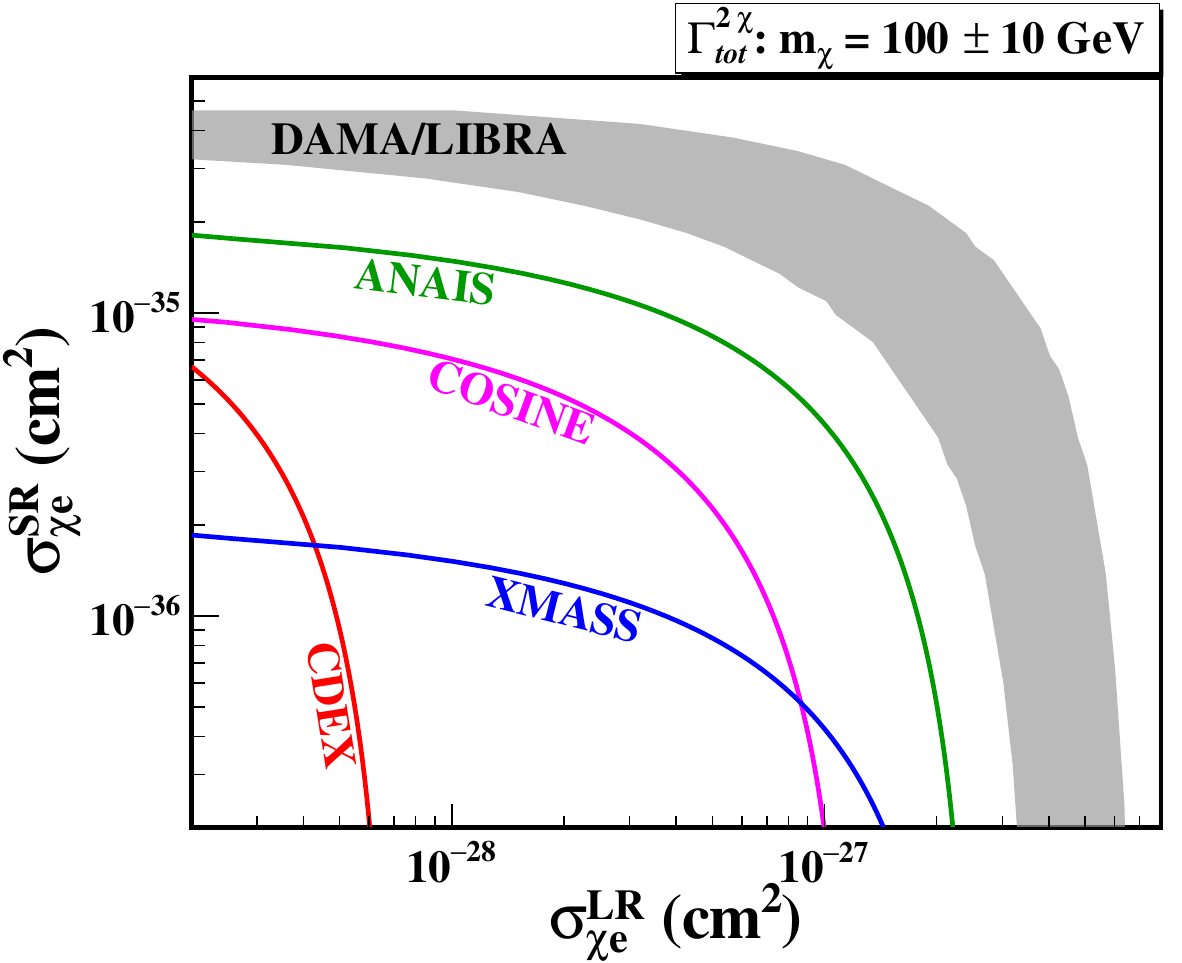}\\
{\bf (c)}\\
\includegraphics[width=0.85\linewidth]{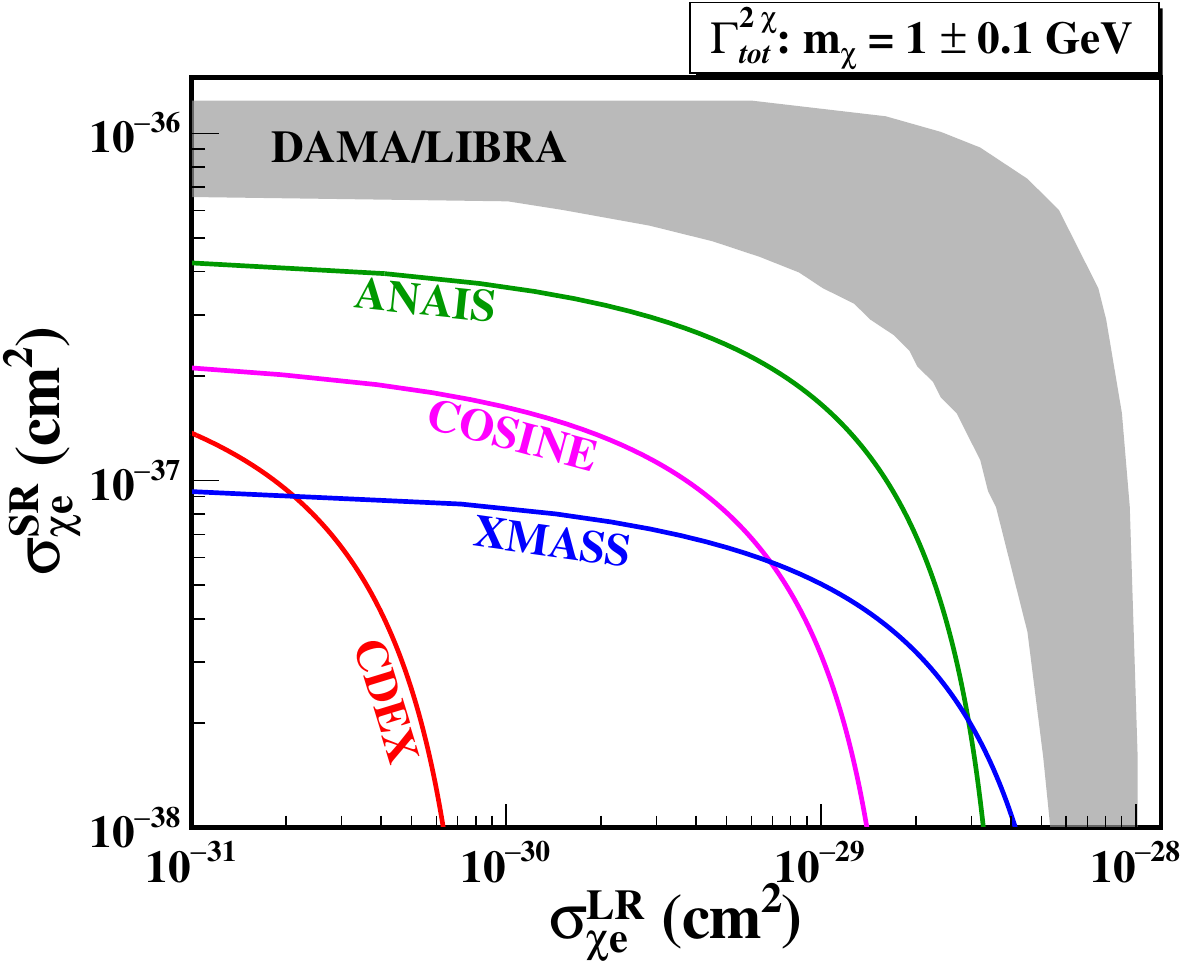}
\end{center}
\caption{
Correlations between  $\SRsigmachie$ and $\LRsigmachie$,
showing the allowed region in the 90\% C.I. from $\DL$ AM data~\cite{dama2019}
and in comparison with the null AM experiments at 90\% C.L. upper limits.
(a)
with the $\1chibf$ analysis, where the best-fit
$\mchi$-range is
$( 60 {\mbox{-}} 100 ) ~ {\rm GeV}$,
and with the $\2chibf$ analysis, at
(b) $\mchi {=} ( 100 {\pm} 10 ) ~ {\rm GeV}$, and
(c)
$\mchi {=} ( 1 {\pm} 0.1 ) ~ {\rm GeV}$.
}
\label{fig::correlation}
\end{figure}

%% =========================

On the contrary, 
no AM signatures were observed in the COSINE, ANAIS, XMASS, CDEX experiments.
Their data are consistent with the null hypothesis. 
As illustrations, the predicted AM spectra in these experiments
due to the best-fit values of ($\SIchiN {,} \LRchie {,} \SRchie$) 
derived from DL data are depicted in Figures~\ref{fig::best-fit-projection}a\&b.
The incompatibility of the DL best-fit values with the null AM experiments 
can be readily visualized.

Conservative and inclusive limits are derived
with single-channel ($\fN$) analysis $-$
$\LRsigmachie$ and $\SRsigmachie$ are set 
to be zero in Eq.~\ref{eq::3d_fit} while
the best-fit values of $\SIsigmachiN$ are evaluated, allowing negative values. 
Limits are derived with the unified approach~\cite{feldman1998}. 
Same procedures apply for $\LRsigmachie$ and $\SRsigmachie$ by assuming 
two other $\sigma$'s are zero.

\subsection{Case of $\1chibf$}

The allowed regions in both $\err3rms$ and 90\% confidence intervals (C.I.) 
for $\SIchiN$, $\LRchie$ and $\SRchie$ as functions of $\mchi$
under $\1chibf$-analysis from $\DL$ AM data~\cite{dama2019}
are shown, respectively, in Figures~\ref{fig::explot}a,b\&c.
The exclusion contours showing upper limits at 90\% confidence level (C.L.)
from the null AM experiments
%% COSINE, ANAIS, XMASS and CDEX experiments 
are superimposed for comparisons in the same plots.
Results of $\fN$ are superimposed 
in Figure~\ref{fig::explot}a for comparison.
The relevant ranges of 
the Earth attenuation and scattering effects correspond to
much higher cross-sections 
and the bounds are outside to the plots displayed.
For instance for the CDEX experiment at 2.4~km of rock overburden,
the upper limits of the sensitivity regions are
$10^{\mbox{-}30}~{\rm cm}^{2}$ for $\SIsigmachiN$~\cite{CDEX:2021cll},
$10^{\mbox{-}24}~{\rm cm}^{2}$ for $\LRsigmachie$, and
$10^{\mbox{-}29}~{\rm cm}^{2}$ for $\SRsigmachie$.

%% =========================

There are two allowed regions in $\fN$ at 90\% C.I. at low and high $\mchi$
corresponding to Na and I-recoils, respectively.
With $\LRchie$ and $\SRchie$ are added 
in $\1chibf$, only the high $\mchi$ region due to I-recoils
is allowed at the same significance.
The best-fit solution of $\mchi$ is shifted 
from $ ( 54.25 {\pm} 4.25 ) ~ {\rm GeV}$ in $\fN$
to
$( 83.3 {\pm} 25.65 ) ~ {\rm GeV}$ in $\1chibf$.

The $\chie$-channels dominate the near-threshold behavior, 
while the high energy (${>} 3 ~ \keVee$) spectra are defined  by $\chiN$.
The two $\chie$ channels are highly correlated 
and share the strength of the AM counts at low energy.
The limiting cases of the best-fit values in $\1chibf$ 
of  $\LRsigmachie$ at $\SRsigmachie {=} 0$ 
and $\SRsigmachie$ at $\LRsigmachie {=} 0$ 
correspond to the light red circles in Figures~\ref{fig::explot}b\&c, respectively.
The exclusion contours of the null AM experiments 
are superimposed.
The weaker bounds in $\LRsigmachie$ in Figure~\ref{fig::explot}b
originates from its sharper rise of the AM spectra
due to an additional $( 1 / q^2 )$ dependence, so that
only the low threshold data from CDEX places strong constraints.

Under $\1chibf$, the DL AM allowed regions in 
Figures~\ref{fig::explot}a,b\&c are inter-dependent $-$
$\mchi$ is constrained to $( 60 {\mbox{-}} 100 ) ~{\rm GeV}$
by the $\SIchiN$ channel.
Figures~\ref{fig::explot}b(c) shows that 
the low cross-section portion of
$\LRsigmachie$($\SRsigmachie$) allowed regions 
at this $\mchi$ range remains unprobed by its own interactions.
However, this parameter space is correlated with the high cross-section portion
of its counterparts.
The correlations are depicted in Figure~\ref{fig::correlation}a
which shows the combined constraints of  $( \LRsigmachie , \SRsigmachie )$ 
from the null AM experiments
can reject the entire DL allowed region in $\1chibf$.
The exclusion margins are particularly large
through the combined constraints of CDEX and XMASS.

%% ===========  chi-e only ==========

\subsection{Case of $\2chibf$}

In the scenario of $\2chibf$ where $\chiN$ and $\chie$ interactions
are due to two different $\chi$'s, 
the constraints due to the high and low energy %%(above and below 3~$\keVee$) 
AM spectral components are independent. 
The $\DL$ $\SIchiN$ allowed regions are defined
by the high-energy component and remain those 
of Figure~\ref{fig::explot}a,
which are well-excluded by the other null experiments.
The AM low-energy component is dominated by the $\chie$ channels
of a different $\chi$ unconstrained in $\mchi$.
This gives rise to allowed regions in $\2chibf$ represented by the light blue bands 
in Figures~\ref{fig::explot}b\&c
for $\LRsigmachie$ and $\SRsigmachie$, respectively.
The $\2chibf$ scenario is excluded by the null AM experiments, independently in 
$\LRsigmachie$ and $\SRsigmachie$.
The $\DL$ 90\% allowed C.I.  in
typical low and high $\mchi$ ranges 
$\left[ \mchi {=} ( 100 {\pm} 10 ) ~ {\rm and} ~ ( 1 {\pm} 0.1 ) ~ {\rm GeV} \right]$
are depicted in Figures~\ref{fig::correlation}b\&c, respectively,
superimposed with the exclusion contours of the null AM data.
The complementary roles of the experiments in probing the
parameter space can be seen.

%% ==================

\section{Conclusion}

Our $\2chibf$ analysis and results indicate that
the DL AM allowed regions associated with $\chie$ are probed
and rejected by the various null AM experiments.
The CDEX~\cite{cdexam2019} data placed more severe constraints on
$\LRsigmachie$ where the $( 1 / q^2 )$ dependence favors
low threshold experiments.
The XMASS~\cite{xmass2018}  experiment is more sensitive
to $\SRsigmachie$ due to its large exposure. 

It can be projected that the DL allowed regions
in generic two-WIMP models 
interacting via the $\chiN$ and $\chie$ channels independently 
are ruled out by the null AM experiments.
Very little room is left to account for the DL AM data with
WIMP-induced $\chiN {+} \chie$ effects.

%% ================================

\section{Acknowledgments}

This work is supported by
contracts 106-2923-M-001-006-MY5,
110-2112-M-001-029-MY3, and
113-2112-M-001-053-MY3
from the National Science and Technology Council, Taiwan,
grants 2021-22/TG2.1 from
the National Center of Theoretical Sciences, Taiwan.

%%  ===  References =========

\bibliography{gdmam.bib}

%% ================================

\end{document}